# Canonical Descriptors for Periodic Lattice Truss Materials

Ge QI[a*], Huai-Liang Zheng[b], Chen-xi Liu[c], Li MA[d], Kai-Uwe Schröder[e]

[a] *Faculty of Civil Engineering and Mechanics, Jiangsu University, Zhenjiang 212013, PR China*

[b] *College of Mechanical and Electrical Engineering, Harbin Engineering University, Harbin 150001, PR China*

[c] *School of Mechanical Engineering, Jiangsu University, Zhenjiang 212013, PR China*

[d] *Center for Composite Materials, Harbin Institute of Technology, Harbin 150001, PR China*

[e] *Institute of Structural Mechanics and Lightweight Design, RWTH Aachen University, Aachen 52062, Germany*

**Abstract:**

**For decades, aspects of the topological architecture, and of the mechanical as well as other physical behaviors of periodic lattice truss materials (PLTMs) have been massively studied. Their approximate infinite design space presents a double-edged sword, implying on one hand dramatic designability in fulfilling the requirement of various performance, but on the other hand unexpected intractability in determining the best candidate with tailoring properties. In recent years, the development of additive manufacturing and artificial intelligence spurs an explosion in the methods exploring the design space and searching its boundaries. However, regrettably, a normative description with sufficient information of PLTMs applying to machine learning has not yet been constructed, which confines the inverse design to some discrete and small scrutinized space. In the current paper, we develop a system of canonical descriptors for PLTMs, encoding not only the geometrical configurations but also mechanical properties into matrix forms to establish good quantitative correlations between structures**

---

* Corresponding author, E-mail address: qige@ujs.edu.cn (Ge QI)

**and mechanical behaviors. The system mainly consists of the geometry matrix for the lattice node configuration, density, stretching and bending stiffness matrices for the lattice strut properties, as well as packing matrix for the principal periodic orientation. All these matrices are theoretically derived based on the intrinsic nature of PLTMs, leading to concise descriptions and sufficient information. The characteristics, including the completeness and uniqueness, of the descriptors are analyzed. In addition, we discuss how the current system of descriptors can be applied to the database construction and material discovery, and indicate the possible open problems.**

## 1. Introduction

Periodic lattice truss materials (PLTMs) stun material scientists, physicists and engineers in recent decades, due to their tremendous diversity of both the topological architectures and physical behaviors[1–4]. Aiming at superior load bearing capability, intensive efforts have been devoted to approach the theoretical upper bound of bulk stiffness and strength[5–9] by designing the material as a stretching dominated architecture[10] and the displacement distribution as an affine-down field[5]. Design strategies concentrating on the lattice struts, including the hollow sections[11–15], the structural hierarchy[16–20] and the bone remodeling struts[21–24], are also implemented to transcend the seminal Gibson-Ashby model. Aiming at counterintuitive performance, negative Poisson's ratio[25–27], negative stiffness[28–30], negative thermal expansion[31,32], etc. are achieved and widely-known as metamaterials or metastructures.

Since the rapid development of additive manufacturing conquers the fabrication challenge of complicated trusses in the last couple years, conventional design concepts, ranging from biomimicry to crystallography, expose their inability to explore the vast untapped design space. Rather than racking human beings' brains, machine learning employs computers to perceive, mine and map between high-throughput input and out data, providing a new promising paradigm to address this difficulty[33–35].

Predicting mechanical properties using surrogate models is being increasingly adapted, with the advantages of higher accuracy and lower computational cost[36–40]. The surrogate models by various machine learning algorithms extract the implicit, profound and complex relations between parameterized geometries/materials and mechanical characteristics. Nevertheless, as a data-driven framework, these relations constructed by computers focuses only on the bare numbers rather than the physical mechanisms, whose promotion of novel lattice truss material discovery is relatively limited. Training the model directly utilizing images[36,38,41,42] or invoking the graph neural networks[33,42,43] elevates the consistency between the inputs/outputs and the realistic structures, whereas, such information supposed to be understood and interpreted is still a black box for human beings.

The inverse problem, i.e. recover the structural parameters according to the desired performance, is a proverbial one-to-many mapping issue in machine learning. In general, the inverse design of lattice trusses is merely performed in some disparate small domains based on prior physical-based formulas, represented as a parameter optimization issue[44–46]. This approach, however, is infeasible when the topological category of the lattice truss changes, ascribed to the different mechanical criteria. If a canonical system of descriptors can encompass and convert adequate information encoded in the lattice truss material, including both the material properties and the geometries, into a dual-readable format for computers and human beings, like the molecular formulas in chemistry, a continuous search space, containing the fragmented well-studied and the immense unexploited domains, will be unified, which can be traversed by advance machine learning algorithms.

With the increasing recognition of the importance of descriptors in machine learning based studies on lattice trusses, some articles have appeared on this topic in recent years. A heuristic language system for PLTMs is laid out by Zok et al.[47], which is employed by Bastek et al.[48] for inverting the structure-property map, due to its generalization and conciseness. On the basis of three fundamental cubic trusses, a series of lexicographic strings is compiled in this concept, yielding a hierarchical classification system from elementary cubic to complex lattices. Although the developed string representations can cover a broad range of tapped lattice truss materials, there still exist three limitations. First, this taxonomy derives all types of lattices from three elementary cubic trusses, simple cubic (SC), body-centered cubic (BCC) and face-centered cubic (FCC) cells. These terms invoke the crystallographic description in the field of metals, which is understandable by human beings, yet is not interpretable by machines. Second, this language denotes only the topological configurations, i.e. “shapes”, of lattice trusses, but assigns no indication regarding geometric levels or material properties. Additional information is required to quantitatively obtain the mechanical behaviors of the lattices. Third, complicated truss cells with multiple nodes and struts will lead to lengthy string descriptions, which may not be readily understood, like the super-long words in the dictionary. Similar merits and limitations are also embodied in the mathematical representations of universal physical networks[49–51].

Parameterizing a series of basic building blocks and hybridizing them are demonstrated to be a exercisable technique to establish a finite search space covering a substantial amount of topologies. This method is applied in the literature[52–58] generating distinct architected lattice trusses in their investigations, respectively, in spite of probably comparable expected modulus or strength. This phenomenon is evidently ascribed to the primitive units, and the diversity of the designed lattice truss structures could be considerable if the huge dataset built by Lumpe and Stankovic[59] is accessed. Using this route, perhaps, one or several species of lattice trusses could be acquired satisfying current engineering demand of mechanical characteristics. Nonetheless, the initial ambition to search the entire design space is not achieved. More recently, according to the common symmetry of lattice cells and the relative positions of the nodes and struts, a vast latent space of 3D truss materials is unified by the adjacency matrix and node position matrix[60]. Herein, the advanced machine learning algorithm is utilized not only to estimate the topology-property interactions, but also to extract the configurational information of the lattice cells. Unfortunately, the base material information is still not embedded within the encoding.

Other investigations attempt bypass the intractability by pixelating the representative unit and converting the configurational information into a binary string or matrix[61–65]. The inherence of pixelation technique makes it more appropriate for 2D material reconstruction rather than 3D structure representation. Since the PLTM with a low relative density consists of spatially arranged slender struts, whose geometric feature is quite difficult to be captured by general resolution. Thus, this method is mainly used for topology optimization of metasurfaces.

Our objective in this article is to propose a canonical description system for the PLTMs, with a dual-readable format for both the human beings and the machines to accelerate the inverse design in the current age of machine learning. This system consists of a series of representative matrices with the same dimensions, which are theoretically derived from the inherent natural of PLTMs, encoding both the geometries (not only the configurations) and the material properties. Through analyzing the completeness and uniqueness, the developed descriptors are demonstrated to provide concise and unambiguous representations for lattice tessellations in 3D space. Furthermore, we show their potential merits in mechanical characterization and inverse design of PLTMs in future studies.

## 2. Desirable features of PLTMs and descriptors

A grand unified description system containing the total information without any tiny omission can undeniably bridge the topology-property gap and unblock the possibility of the forward prediction as well as the inverse design for PLTMs at any scales, but such a representation system is inevitably too complicated according to the "no free lunch" theorem[66]. Consequently, the PLTMs described in the current paper should satisfy some basic requirements to reduce the complexity as follows:

(i) No isolated node: all the lattice nodes within one PLTM are joined by at least one strut and no node is unconnected.

(ii) No isolated strut: all the lattice struts within one PLTM are interconnected at the nodes and no strut exists in isolation.

(iii) No repeated nodes: there can be one and only one lattice node at a precise location in the lattice space.

(iv) No isolated sub-part: the lattice struts and nodes in one bulk of lattices should be interconnected due to the continuum of the materials. This requirement can be deduced by the "no isolated strut" requirement.

(v) No strut intersection: all the struts within one lattice truss structure are interconnected only at the lattice nodes, and no strut intersection occurs at any other positions.

(vi) 3D periodicity: the lattice truss structures comprise repeated arrays of the basic element (i.e. the representative cell) in three orthogonal principal directions.

(vii) Straight strut: all the struts within one lattice truss structure are tacitly straight. The cross-section can be circular or polygonal. Curved strut or variable-cross-section struts are not covered in current study.

Note that the first six requirements are essential to constitute a bulk PLTM, while the straight strut principle is only adapted in the current article which does not cover the topologies adopting curved elements[67]. Similarly, some basic requirements of lattice descriptors are shown as

(i) Structural interpretation: the descriptors in any forms (numbers, mathematical symbols, languages and graphs) cannot be arcane. They must be logical, meaningful and rigorous representation, which is capable of conveying information. The encoding or describing rules can be rationally interpreted.

(ii) Show good correlation with mechanical properties: the mechanical properties of the lattice truss structure can be sized by the lattice descriptor with some essential procedures.

(iii) Generalization ability: the proposed lattice descriptors can not only describe the examples in this paper, but also carry over into more categories of lattices by rational transformation.

(iv) Simplicity: complexity of the lattice descriptor should be minimized for the feasibility.

(v) Uniqueness: one-to-one mapping should be guaranteed between the lattice descriptor and the lattice truss structure, exactly as the fingerprint.

(vi) Completeness: no indispensable information loss occurs during the encoding of the lattice descriptor.

## 3. Creation and development of descriptors

By their very nature, PLTMs are constitutionally defined by a set of nodes and struts. The periodic arranged struts comprise the physical entity of the PLTMs, while the nodes denote the endpoints and the interconnection positions of struts. This scenario just coincides with the molecular structures, including the atoms and the covalent bonds [68]. Inspired by this, matrices will most likely be the mathematical candidate notations ascribed to their advantaged graph-theoretical representation.

### 3.1. Representation by matrices

First of all, on the basis of the possible lengths of struts, i.e. the relative geometrical distances between the lattice nodes in 3D space, we develop the *geometry matrix*, denoted by $\mathbf{G}$, as

$$\mathbf{G} = \begin{bmatrix} 0 & r_{12} & \cdots & r_{1n} \\ r_{21} & 0 & \cdots & r_{2n} \\ \cdots & \cdots & \cdots & \cdots \\ r_{n1} & r_{n2} & \cdots & 0 \end{bmatrix} \tag{1}$$

where $n$ is the number of the lattice nodes in one represented lattice cell and the entry $r_{ij}$ denotes the Euclidean distance between *Node i* and *Node j*. According to the inherent nature of its definition, $\mathbf{G}$ is a hollow square symmetric matrix with $n$ rows and columns.

It is evident that the geometry matrix only involves the knowledge of the lattice node configurations (shape) and geometries (size), but does not contain sufficient information about the struts. Thus, it is accompanied by a series of matrices to define the representative cell. The *density matrix*, denoted by $\mathbf{D}$, is proposed to complement the mass information of each lattice struts:

$$\mathbf{D}=\begin{bmatrix} 0 & \lambda_{12} & \cdots & \lambda_{1n} \\ \lambda_{21} & 0 & \cdots & \lambda_{2n} \\ \cdots & \cdots & \cdots & \cdots \\ \lambda_{n1} & \lambda_{n2} & \cdots & 0 \end{bmatrix} \tag{2}$$

where the entry $\lambda_{ij}$ is the linear density of each strut. Besides, we employ *stretching stiffness matrix* $\mathbf{K}_t$ and *bending stiffness matrix* $\mathbf{K}_b$ for the mechanical responses of the lattice struts:

$$\mathbf{K}_t=\begin{bmatrix} 0 & t_{12} & \cdots & t_{1n} \\ t_{21} & 0 & \cdots & t_{2n} \\ \cdots & \cdots & \cdots & \cdots \\ t_{n1} & t_{n2} & \cdots & 0 \end{bmatrix} \tag{3}$$

$$\mathbf{K}_b=\begin{bmatrix} 0 & b_{12} & \cdots & b_{1n} \\ b_{21} & 0 & \cdots & b_{2n} \\ \cdots & \cdots & \cdots & \cdots \\ b_{n1} & b_{n2} & \cdots & 0 \end{bmatrix} \tag{4}$$

In these two equations, the entries $t_{ij}$ and $b_{ij}$ denote the stretching (or compressive) and bending stiffness of each lattice strut connecting *Node i* and *Node j*, respectively. And they can be calculated by $t_{ij}=E_{ij}A_{ij}$ and $b_{ij}=E_{ij}I_{ij}$, where $E, A, I$ are the Young's modulus of the parent material, the cross-sectional area and the moment of inertia of the strut, respectively. The density matrix, stretching stiffness matrix and bending stiffness matrix are also hollow symmetric matrices with the size of $n \times n$. If *Node i* and *Node j* are connected by one strut, the entries would be positive, while if no any strut exists between two nodes, the values would be zeros.

At this point, a single representative cell of lattice truss materials can be identified by virtue of the above four descriptors, including the configuration, the conformation and the mechanical properties. Then, the information regarding the periodic tessellation of this cell should be revealed, leading to the *packing matrix* $\mathbf{P}$:

$$\mathbf{P} = \begin{bmatrix} 0 & p_{12} & \cdots & p_{1n} \\ p_{21} & 0 & \cdots & p_{2n} \\ \cdots & \cdots & \cdots & \cdots \\ p_{n1} & p_{n2} & \cdots & 0 \end{bmatrix} \tag{5}$$

Differing the continuous parameters in aforementioned descriptors, the candidate values for the entry $p_{ij}$ are discrete in current the packing matrix. Consider a ray with origin *Node i* and passes through *Node j*. Non-zero values are assigned to $p_{ij}$ only if the ray is parallel to one of the principal axes and the two lattice nodes locate on the external surface (including the edges and vertices) of the *oriented bounding box* (OBB) of the *unit cell* (UC). More specifically, if the ray and the principal axis $u_d$ ($d$=1,2,3) are parallel and are in the same direction, $p_{ij} = d$ , while if the ray and the principal axis $u_d$ ($d$=1,2,3) are parallel but are in the opposite direction, $p_{ij} = -d$ . Otherwise, the entries of the current matrix are specified as zero. At this point, the packing matrix presents a hollow skew symmetric matrix.

The packing matrix naturally clarifies the principal directions and the periodic arrangements of the PLTMs. If there is one, two or three different non-zero values in the packing matrix, it allows the UC to be tiled into 1D, 2D and 3D arrays respectively. In conjunction with the geometry matrix, the size of the UC can also be computed. Note that the non-zero values in the packing matrix can be exchanged, but they are recommended to be in the right-hand screw sense, satisfying the canonical coordinating introduced in next section.

The illustrative examples of three ad-hoc lattice trusses are depicted in **Fig. 1**. According to a reference polyhedron or a reference crystal, they are identified as so-called simple cubic (SC), body-centered cubic (BCC) and face-centered cubic (FCC) lattices for the architectural topologies. By the definition of Zok, et al.[47], the 2×2×2 tessellations in three principal directions shown in **Fig. 1A** should be denoted $\{2SC\}^3$, $\{2BCC\}^3$ and $\{2FCC\}^3$, encoding information of the cell type, the number of cells and the number of directions in which the trusses are tiled in space. By our definition, if the conformation of theses UCs are assumed cubes with a side of length 2a (**Fig. 1B**), the Euclidean distance between any two lattice nodes can be obtained, contributing to the geometry matrix. In the case that all the lattice struts are slender cylinders with the section radius of *R*, the linear density, stretching stiffness and bending stiffness of struts are derived as $\pi\rho_s R^2$, $\pi E_s R^2$ and $\pi E_s R^4/4$, yielding the density matrix, stretching stiffness matrix and bending stiffness matrix. Herein, $\rho_s$ and $E_s$ denote the volumetric density and Young's modulus of the parent material. These descriptors for the SC lattice in **Fig. 1A** are simplified shown in **Fig. 1C**, and the complete versions for SC, BCC and FCC lattices are summarized in **Supplementary Table 1~3**.

It is worth noting that when the value of 1 is assigned to all the non-zero entries of the density matrix, stretching stiffness matrix and bending stiffness matrix, these descriptors will degenerate into the widely known *adjacency matrix* $\mathbf{A}$, which contains information regarding lattice node connectivity. Combining the adjacency matrix with the geometry matrix and packing matrix, the descriptors would focus on the topology, geometry as well as the 3D layout of the UCs, without information about functionalities. Moreover, the entries in the density matrix, stretching stiffness matrix and bending stiffness matrix, can be tailored to various requirements, as long as they are encoding information of the struts. For example, the relevant contents could be substituted by the diameters of the cross-sections, the volumetric density, the Young's modulus, the damping loss factor, the thermal expansivity, etc. of the parent materials.

### 3.2. Independent variables

Since the developed descriptor system for PLTMs consists of five hollow square matrices, each of which contains $n^2$ entries, the total amount of entries is therefore $5n^2$. However, some of them are dependent parameters and can be derived from other variables in a natural way. Finding the number of the independent variables of the descriptors would efficiently reduce the computational costs in generation of descriptors and reconstruction of PLTMs.

It is a remarkable fact that, the independent parameters in the stretching stiffness, the bending stiffness and the density matrices are equal, in the sense that knowledge of one implies knowledge of the others. As a hollow symmetric matrix, the theoretical number of independent entries in the density matrix is $n(n-1)/2$, at most. By the same token, an $n \times n$ packing matrix has no more than $n(n-1)/2$ independent entries.

The only exception is the geometry matrix, possessing fewer variables to be determined. We can give a sharp upper bound on the number of independent entries in the geometry matrix in this investigation: $4n-10$, and it is not difficult to think of this value. Notice that if $n$ lattice nodes are arranged in 3D space of the UC in general position, i.e. they do not lie on a line or in a plane, a Cartesian coordinate system can be defined via four nodes, which are not coplanar. The distances each pair of these four nodes are indispensable, inducing six independent entries. Based on this coordinate system, every additional lattice node can be orientated, adding four independent values. As to the distances between those additional lattice nodes, the values can be easily computed using the Euclidean distance formula.

### 3.3. Characteristics of the matrices

Since any descriptors, whether expressed by the proposed symbols in matrices or illustrative representations in graphs, indicates the same inherent essence of PLTMs, the characteristics of PLTMs must be revealed by the descriptors in an explicit or implicit way, and vice versa.

The geometry matrix, revealing the node distribution of the PLTMs, could be used to identify the “no repeated nodes” rule in §2. The “no repeated nodes” rule prescribes that a cluster of repeated nodes should be coalesced into a single one, and therefore, the Euclidean distance between any two lattice nodes must be greater than zero. Except for the diagonal entries, all other values in the geometry matrix must be positive (**Fig. 2A**).

For the stretching stiffness matrix, the bending stiffness matrix and the density matrix, they are utilized to identify the “no isolated node”, “no isolated strut” and “no isolated sub-part” principles due to their indications of strut conditions. Considering the same composition form among these three matrices, hereafter, the characteristics are resolved taking merely the density matrix as example.

The “no isolated node” rule prescribes the minimum number of struts connected to one single lattice node. If the entry $\lambda_{ij}$ is zero, *Node i* and *Node j* are not directly connected by any struts. In this interpretation, any all-zero row or all-zero column in $\mathbf{D}$ matrix are not permitted, see **Fig. 2B**.

We adopt an analogous method to evaluate the “no isolated strut” rule. If a strut connecting *Node i* and *Node j* is independently present without any link to other struts, all the entries, except for $\lambda_{ij}$ and $\lambda_{ji}$, in the *i*th and *j*th rows as well as *i*th and *j*th columns are zeros. This condition is absolutely not permitted in the density matrix, and can be visibly identified, as shown in **Fig. 2C**.

The further inference of the “no isolated strut” rule denotes the “no isolated sub-part” principle, with the counter-example shown in **Fig. 2D**. We can start the identification process of isolated sub-parts from searching the first row in the density matrix, corresponding to *Node* 1. According to “no isolated node” rule, there is at least one another node directly connected to *Node* 1, just as *Node* 5, corresponding to the non-zero entry $\lambda_{15}$. Subsequently, the fifth column, or the fifth row, is detected to find the non-zero value except for *Node* 1. By analogy, the matrix is searched in turn by the rows and columns, using the indices corresponding to the non-zero entries, until no new row or column joins. All of the nodes should, probably more than once, appear in this network for an intact lattice cell. On the contrary, if some lattice nodes are omitted in this network, the structural defect of isolated sub-parts exist.

As for the packing matrix, it mainly presents the periodic tessellated conditions of repeated UCs in space. The PLTMs in this paper can both denote "structures" on the macroscopic or the mesoscopic scales, and "materials" on the microscopic scale. Hence, the UC naturally provides the ability to be stacked into 1D, 2D or 3D arrays, stipulating the minimum quantity of non-zero integers in the packing matrix. A UC being able to be joined with one neighboring cell means that at least one pair of lattice nodes exists at the symmetric positions on the opposite surfaces normal to the packing direction. Thereupon, at least two entries $p_{ij}$ and $p_{ji}$ in the packing matrix have the non-zero value, if the two nodes are labelled as *Node i* and *Node j*. The minimum quantities of non-zero entries in the packing matrix of a UC, should be, by analogy, four and six for 2D and 3D tessellating ability. And a UC with the packing matrix whose entries are all zeros cannot be periodically arranged, as displayed in **Fig. 2E**.

In addition, though the possible defect of the strut intersection is not explicitly reflected by the descriptors, it can be identified via calculating the spatial relationships of the lattice struts. When the UC is coordinated, this problem could be rectified via a simple algorithm, which is offered in the **Supplementary materials**.

## 4. The canonical orderings of lattice nodes

If a certain PLTM is given, with information ranging from lattice node positions to strut properties, we can precisely determine the entries of the matrix descriptors in §3, say, by definition. Nevertheless, the index, i.e. the row and the column in which one entry appears cannot be determined until the sorting of the lattice nodes is articulated. On the other hand, if the lattice nodes are sorted in an arbitrary way, this would result in, in theory, $n!$ possible groups of descriptors. Consequently, it is vital to distinguish a finite number of orderings of the lattice nodes, that turn out to be invariant when an orientation-preserving affine transformation is performed. These orderings intrinsic to the UC itself are defined as the *canonical orderings* associated to the PLTM.

### 4.1. Canonical coordinatizing and sorting

If the space is coordinatized, these lattice nodes can be lexicographically ordered by the coordinates. Due to the periodic packing feature, the OBB of the UC of a PLTM denotes a cuboid (more often a cube), an architecture shape to which the Cartesian coordinate system can applied in a natural way. In this interpretation, any of the vertices and the corresponding mutually perpendicular edges of the cuboid could be used as the origin and as the axes of the Cartesian system, respectively. Assuming the three axes form a right-handed system and the UC lies in the positive side of each axis, the number of possible coordinate systems can be determined as 24. It is derived that the OBB yields eight alternative positions for the origin, where each origin possesses three alternative sets of axes. Then, the lattice nodes are spontaneously sorted by firstly $u_1$-, then by $u_2$- and finally by $u_3$-coordinates in ascending order, with three examples displayed in **Fig. 3A** when the origin location is fixed. The all 24 canonical orderings of the DC UC are presented in **Supplementary Figure 2**.

It is now easy to realize that the canonical orderings present more concise for 2D PLTMs. The method, as already suggested, is to locate a right-handed coordinate system to one vertex of the OBB. This would contribute to four distinct order types and an example for a planar lattice configuration is schematically depicted in **Supplementary Figure 3**.

**4.2. Mutual transformations of canonical orderings**

If two PLTMs with the equal nodes are given, and we wish to identify whether they are the same, we need only compare their canonical descriptors. However, suppose the giving orderings of the lattice nodes are different, the matrices would be superficially distinct. It would seem that we have to list all the canonical orderings of one PLTM, to be sure of finding all those (if any) in which the canonical descriptor combination would be consistent with that of the other PLTM. Starting from the initial step of canonical coordinatizing to the step of calculating each entry of the matrices appears inefficient, since different orderings change just the indices of the entries rather than the values. Notice that the canonical coordinate systems for a UC of a PLTM remain the right-handed form despite various positions of the origins. It means that the coordinates of all the lattice nodes can be acquired by solid-body transformations, and then the nodes can be sorted.

If the UC of a given PLTM has been canonically coordinatized and the *homogeneous coordinates* of all the nodes are $N_i = (x_{i1}, x_{i2}, x_{i3}, 1)$, the coordinates corresponding to the other two coordinate systems with the same origin should be $N_i' = (x_{i3}, x_{i1}, x_{i2}, 1)$ and $N_i'' = (x_{i2}, x_{i3}, x_{i1}, 1)$, through directly exchanging the coordinates. Another approach is applying rotations as $N_i'^T = \mathbf{R}_o N_i^T$ and $N_i'^T = \mathbf{R}_o N_i'^T$, and the rotation matrix is

$$\mathbf{R}_o = \begin{bmatrix} 0 & 0 & 1 & 0 \\ 1 & 0 & 0 & 0 \\ 0 & 1 & 0 & 0 \\ 0 & 0 & 0 & 1 \end{bmatrix} \quad (6)$$

With this matrix, the canonical orderings in every row of **Supplementary Figure 2** can be transformed into each other, as exhibited in **Fig. 3A**. Then, the case that the origin of the coordinate system moves to another vertex, i.e. the coordinate transformation in each column, would be resolved via a rotation and a translation, yielding the matrices as

$$\mathbf{TR}_1 = \begin{bmatrix} 0 & 1 & 0 & 0 \\ -1 & 0 & 0 & l_1 \\ 0 & 0 & 1 & 0 \\ 0 & 0 & 0 & 1 \end{bmatrix} \quad (7)$$

$$\mathbf{TR}_3 = \begin{bmatrix} 0 & 0 & -1 & l_3 \\ 0 & 1 & 0 & 0 \\ 1 & 0 & 0 & 0 \\ 0 & 0 & 0 & 1 \end{bmatrix} \quad (8)$$

where $l_1$ and $l_3$ denote the size of the UC in $u_1$- and $u_3$-directions of the initial coordinate system. $\mathbf{TR}_1$ can facilitate the coordinate computation when the origin moves in $u_1u_2$-plane, as shown in **Fig. 3B**, addressing coordinate reassignment of the first or the last four rows in **Supplementary Figure 2**. In a similar way, the transforming from the first row to the fifth row is realized by $\mathbf{TR}_3$. Consequently, beginning from any one canonical order with the coordinate system, we would get all the 24 canonical orderings.

### 4.3. Transforming random numbered descriptors into canonical prototype

The entries of the descriptors are, sometimes, calculated with a random numbered lattice node sequence. The current problem is to find all permutations such that the descriptors are presented in the canonical prototype employing the existing results, instead of reproducing all the matrices.

**Fig. 3C** shows two UCs that clearly belong to the same PLTM, but not in the numbering shown. If the geometry matrix $\mathbf{G}$ is induced by the left random labeled configuration while $\mathbf{G}'$ is generated by the right canonical one. The permutation will have to map $\{1,2,\cdots,13,14\}$ to $\{14,11,\cdots,5,3\}$ in 1-1 correspondence so that the two geometry matrices are in exact consistency. Herein, the permutation is executed by $\mathbf{G}' = \Pi \mathbf{G} \Pi^T$ , where $\Pi$ is the permutation matrix. Its entry $\pi_{ij} = 1$ when the subscript *i* denotes the label of the canonical ordering of the lattice node while *j* denotes that of the random sequence, respectively, and $\pi_{ij} = 0$ otherwise. For example, $\pi(14,1)$, $\pi(11,2)$ and $\pi(3,14)$ should be 1 for the permutation in **Fig. 3C**.

**5. Completeness and uniqueness**

It is a fairly apparent matter to prove the combination of $\mathbf{G}$, $\mathbf{D}$, $\mathbf{K}_t$, $\mathbf{K}_b$ and $\mathbf{P}$ matrices is complete, that is to say, they are capable of describing all the geometric features and mechanical properties of the PLTMs. Meanwhile, the elastic constants or mass information of the PLTMs are also implicit in these descriptors, since the stiffness properties (including tensile and bending) and density distributions for each strut are involved. Homogenization theory is intended to provide an approach to determine the effective elastic moduli [69–71].

This is no longer true, however, for the related problem of uniqueness. The definitions of these matrices indicate that they will vary when the lattice nodes are labeled by a different order, as stated in §4. If, of course, the lattice nodes can be exactly sorted into only one sequence, our descriptors are naturally unique. But this question is still a matter of active interest in applied mathematics, and it has been confirmed that the rigorous uniqueness, i.e. that for a physically given PLTM there exists one and only one group of descriptors, cannot be pursued. The number of the canonical orderings is 24 for 3D structures (4 for 2D configuration), respectively, which presents a finite and fixed quantity.

Nonetheless, when a different canonical sorting is applied to a PLTM, the indices of the entries of the matrices are adjusted, but the values and the corresponding number of the entries remain the same. By rational permutation, one canonical ordering of lattice nodes can be mapped to any other canonical orderings, so that their nodes can be put in 1-1 correspondence. From this interpretation, we can recognize that the 24 sets and 4 sets of canonical descriptors for 3D and 2D PLTMs are equivalent respectively. In a non-rigorous sense, the uniqueness of the developed lattice descriptor is regarded to be achieved.

## 6. Applications of the descriptors

To illustrate the merits of the developed descriptors for PLTMs, we in this section introduce the method of reconstructing a PLTM from the descriptor system. Furthermore, the mechanical properties of the PLTMs directly derived from the descriptors are represented, demonstrating the relevance in the fields of optimization and inverse design.

### 6.1. Reconstruction of a PLTM

Despite the simplicity of the matrix representation of PLTMs, depiction from some certain matrices to 3D structures is not intuitive, since the coordinates are not explicitly listed. The reconstruction route of a PLTM from matrices to 3D graph is schematically illustrated in **Fig. 4**.

Inherently, the reconstruction of a PLTM is determination of the spatial positions of the struts and the nodes relative to the OBB. Therefore, a temporary right-handed coordinate system is initially established. *Node* 1 is assumed to locate at the origin and *Node* 2 lies on the positive direction of the $u_1'$-axis. Subsequently, the first *Node i* which is not on the $u_1'$-axis would lead to the $u_2'$-axis as shown in **Fig. 4A**. Notice that the $u_3'$-coordinate of the first *Node j* which does not lie in the $u_1'u_2'$-plane may be either positive or negative, and the different options will create two mirrored structures, unless the PLTM itself is achiral. Relying on these four nodes, all the rest nodes can be oriented through the geometry matrix. And then the density, stretching and bending stiffness information will produce the lattice struts. Following that, the point is determining orientation of the OBB corresponding to the given descriptors, i.e. that transforming the temporary coordinate system to the canonical prototype. The packing matrix clarifies the directions of the OBB and the canonical coordinating can be realized by transformation. Finally, we compare 48 canonical orderings (including the two mirrored configuration), which are gained using the approach in §4.2 , to the given version, and the exact structure of the PLTM is obtained.

### 6.2. Determination of mechanical properties

Maxwell[72] and Deshpande et al.[10] suggest three key characteristic parameters describing physical frames: the number *s* of lattice nodes, the number *e* of lattice struts and the connectivity *Z* of the PLTMs, whose algebraic relations distinguish the determinateness of the mechanical properties of the PLTMs. For a UC, the determination of these key parameters is reasonably straightforward. The node number, according to the definition, equals to the size of the describing matrices, i.e. $s_{unit} = n$, where the subscript denotes the UC. And the strut number denotes one half of the number of non-zero entries in the density matrix (or the stretching stiffness matrix as well as the bending stiffness matrix), i.e. $e_{unit} = \|\mathbf{D}\|_0 / 2$. Thereupon, the UC takes on a value of $Z_{unit} = 2e/s = \|\mathbf{D}\|_0/n$. For the finite or infinite arrays in 1D, 2D and 3D packings of UCs, the computation of *e*, *s* and *Z* will become more complicated, due to the mergence of nodes and struts at the interfaces. The solution to this challenge will be investigated in the future study.

The weight $W$, density $\rho$ and relative density $\bar{\rho}$ are also important indicators characterizing PLTMs. For a UC, we can calculate the weight as

$$W = \frac{\|\mathbf{G} \odot \mathbf{D}\|_1}{2} \tag{9}$$

Herein, if three principal directions are explicitly represented in the packing matrix, the density of the PLTM is

$$\rho = \frac{\|\mathbf{G} \odot \mathbf{D}\|_1}{2 r_{i_1 j_1} r_{i_2 j_2} r_{i_3 j_3}} \tag{10}$$

in which, the operation "$\odot$" denotes the Hadamard product and the subscripts $i_k j_k$ (*k*=1,2,3) satisfy that the entry $p_{i_k j_k} = \pm k$ in the packing matrix. Then, the relative density should be $\bar{\rho} = \rho/\rho_s$, when all the struts are made by the same parent material.

The elastic properties of PLTMs can be deduced by generating the rigidity relations of each strut between the node loads and deformations. The stretching relation is initially determined by the stretching coefficient matrix

$$\mathbf{C}^{et} = \mathbf{K}_t \oslash \mathbf{G} \tag{11}$$

where the operation "$\oslash$" denotes the Hadamard division, i.e. $C_{ij}^{et} = t_{ij}/r_{ij}$. Particular attention should be paid to the diagonal entries $C_{ii}^{et} = \omega$ when the divisor $r_{ii} = 0$. Analogously, the relationships between imposed forces/moments and the flexural deformation (including the rotation and the deflection) of the lattice nodes are determined by the bending coefficient matrices

$$\begin{cases} \mathbf{C}_1^{ef} = \mathbf{K}_b \oslash \mathbf{G} \\ \mathbf{C}_2^{ef} = \mathbf{K}_b \oslash (\mathbf{G} \odot \mathbf{G}) \\ \mathbf{C}_3^{ef} = \mathbf{K}_b \oslash (\mathbf{G} \odot \mathbf{G} \odot \mathbf{G}) \end{cases} \tag{12}$$

The diagonal entries of these three matrices are also "undefined" with the value of $\omega$. A group of the entries with the same indices in these four matrices encodes the rigidness information for one strut if it is regarded as a 3D slender beam, and in this interpretation, the diagonal entries mean that a lattice node itself cannot deform in physicality. We can derive the elastic constants of PLTMs using the direct stiffness method in finite element analysis (FEA). Some details are introduced in the **Supplementary materials**, while the complete computing process is omitted in current study.

## 7. Concluding remarks and open problems

How many distinguish PLTMs are there and how are the mechanical behaviors of them? This is evidently a difficult question, but we can compile a "dictionary" utilizing our system of canonical descriptors for PLTMs. This system employs matrices to describe the intrinsic information of lattice nodes and struts in a concise yet unambiguous way. So the structural characteristics as well as possible defects can be identified by the entries. The more pertinent point is that the developed descriptors can realize the reconstruction and determination of mechanical properties of PLTMs, implying that the studies on quantitative structure-property relationships appear to be encouraging.

Since the canonical descriptors are represented by matrices in mathematics instead of words in a spoken language, it become possible to use a computer to yield many descriptors, some possible applications and problems immediately suggest themselves. Two among them are:

(a) *To database construction*. Numerous researches [54,58–60,73] demonstrate that the configuration, i.e. the architecture topology, of the UC is the most vital factor affecting its mechanical responses. The descripting system we propose in present paper simultaneously compiles the configuration, the conformation and the mechanical properties of PLTMs. Consequently, performing a clustering analysis on the matrices, one can group structurally similar PLTMs together, and build a so-called standard database. The "look-up" of representatives from each cluster produced from a set of PLTMs should be sufficient to understand the structure-property relationships of the whole set, without the need to test them all. Alternatively, if the topological configuration of the UC is too complicated to make such a "look-up" computationally feasible, one can abstract characteristics of the standard matrices, and check off the corresponding characteristics of the new descriptors against them.

There is probably a phenomenon in this application that the clusters are probably set according to the dimension of the matrices, i.e. the number of lattice nodes, ignoring the mechanical mechanisms. So it suggests a problem how to find a fast algorithm for classifying PLTMs based on the inherent mechanisms instead of the superficial lattice nodes.

(b) *To material discovery*. Discovering genuine novelty in PLTMs is our most desired purpose. Now that our developed descriptors are cognizant to both humans and computers, inverse design of PLTMs with tailored properties can be resolved using advanced machine learning methods. A preferred route is introduced as follows: initially, yield the mechanical responses of described PLTMs by FEA; subsequently, build an inverse model to learn the correlation between the performance and the descriptors; finally, generate new descriptors through the inverse model and reconstruct the UC.

A challenge in this strategy denotes that the descriptors proposed in the present investigation possess different dimensions, which is a technical difficulty for machine learning. So it suggests a problem how to find an approach for transferring the descriptors into a physically interpreted form with the same and lower dimension.

**Acknowledgements**

This present work is supported by the National Nature Science Foundation of China under Grant No. 12202161, the Natural Science Foundation of Jiangsu Province under Grant No. BK20220512 and the Natural Science Foundation of the Jiangsu Higher Education Institutions of China under Grant No. 22KJB130002.

**Fig. 1. Schematic illustration of a series of ad-hoc periodic lattice truss materials. A**, The three lattice truss materials, whose topological configurations are loosely described by the terminologies of simple cubic (SC), body-centered cubic (BCC) and face-centered cubic (FCC) in prior research, comprise a finite packing tessellation with 2×2×2 arrays of the unit cells in 3D space. The white spheres and blue bars indicate the lattice nodes and struts, respectively. **B**, The OBB of the unit cell of these lattices are assumed cubes with a side of length $2a$ and their nodes are randomly numbered. All the struts of these lattice trusses are assumed to have the same cross-section, which is a circular with the diameter of $d$. **C**, The descriptors for the SC (**A**) are calculated based on the geometry and material information. Note that the "Symmetry" indicates that the matrices are symmetric and "Skew symmetry" means that the matrix is skew symmetric and the corresponding entries are omitted. The matrices corresponding to the strut distribution, i.e., $\mathbf{D}$, $\mathbf{K}_t$ and $\mathbf{K}_b$, present the same characteristics of entry distribution except the coefficients, thus allowing the simplified representation.

**Fig. 2. The counter-example graphs of the expected characteristics of the PLTMs and the corresponding explicit representations in the matrices. A**, The counter-example to the "No repeated nodes" rule. There are repeated points at one vertex of the unit cell, i.e., *Node* 5 and *Node* 6, between which, the Euclidean distance $r_{56}$ and $r_{65}$ is zero. Correspondingly, two anomalous zeros appear in the geometry matrix. Note that the repeated points are not placed at exactly the same coordinate for purposes of illustration. **B**, The counter-example to the "No isolated node" rule. An isolated lattice node (*Node* 5) is inserted into a SC cell, generating an all-zero row and column in the density matrix, as marked in red. **C**, The counter-example to the "No isolated strut" rule. The orange strut connecting *Node* 3 and *Node* 5 losses all its links to the rest part of the lattice, causing that all the entries except for $\lambda_{35}$ and $\lambda_{53}$ in two corresponding pairs of rows and columns are zero. **D**, The counter-example to the "No isolated sub-part" rule. In this unit cell, the two sub-parts are formed by placing struts along three non-coplanar edges a tetrahedron (the green part), and by placing struts normal to and at the center of each face of a tetrahedron (the orange part), respectively. From the view of the non-zero entries in the density matrix, *Node* 1, 3, 5, 7 and 8 form a network, in which, *Node* 2, 4, 6 and 9 are missed. Thus, the unit cell is not an intact bulk and violates the "No isolated sub-part" principle. **E**, The counter-example to the "3D periodicity" rule. The tetrahedral

unit cell arranged in this unit cell could not be tightly tessellated into 1D, 2D or 3D arrays without any gap along the current principal directions, leading to an all-zero packing matrix.

**Fig. 3. The canonical orderings of a lattice unit cell based on an illustrative diagram of the diamond cubic (DC) truss. A**, The mutual transformations of the canonical orderings corresponding to three coordinate systems with the same origin. For instance, three groups of possible Cartesian systems, indicated in red, green and blue respectively, might be applied to the DC when the origin is located at the vertex *o*. The coordinates of the lattice node are computed via the rotation matrix and the canonical orderings are immediately gained. **B**, If the origin of the coordinate system moves to the vertex on the $u_1$-axis, the new nodes coordinates would be calculated by a rotation and a translation, which can form a translation-rotation matrix $\mathbf{TR}_1$. **C**, For a PLTM with random numbered nodes, a permutation $\pi$ can map the random sequence to the canonical sort and the matrices can be computed in a simple way, instead of starting all over again.

**Fig. 4. Schematic overview of the reconstruction procedure of a PLTM from descriptors. A**, A Cartesian coordinate system is establishes based on the first four non-coplanar lattice nodes. During this step the *Node* 1 and *Node* 2 are used to set the temporary $u_1'$-axis, and the first non-collinear *Node i* is set in the positive direction of the temporary $u_2'$-axis. Then, the right-handed coordinate system can be determined. Two possible positions in the mirror symmetry are present for the first non-coplanar node, perhaps, leading to two mirrored structures. **B**, The coordinates of all the rest lattice nodes can be accurately computed based on the entries in the geometry matrix, followed by generation of lattice struts based on the density matrix, stretching and bending stiffness matrices. **C** and **D** display an example of the reconstruction from given matrices to the structure of a pyramidal unit cell. The length of the struts is assumed to be *l*. The temporary coordinate system might not be the canonical prototype and thus the coordinate transformation should be applied to be consistent with the originally defined one.

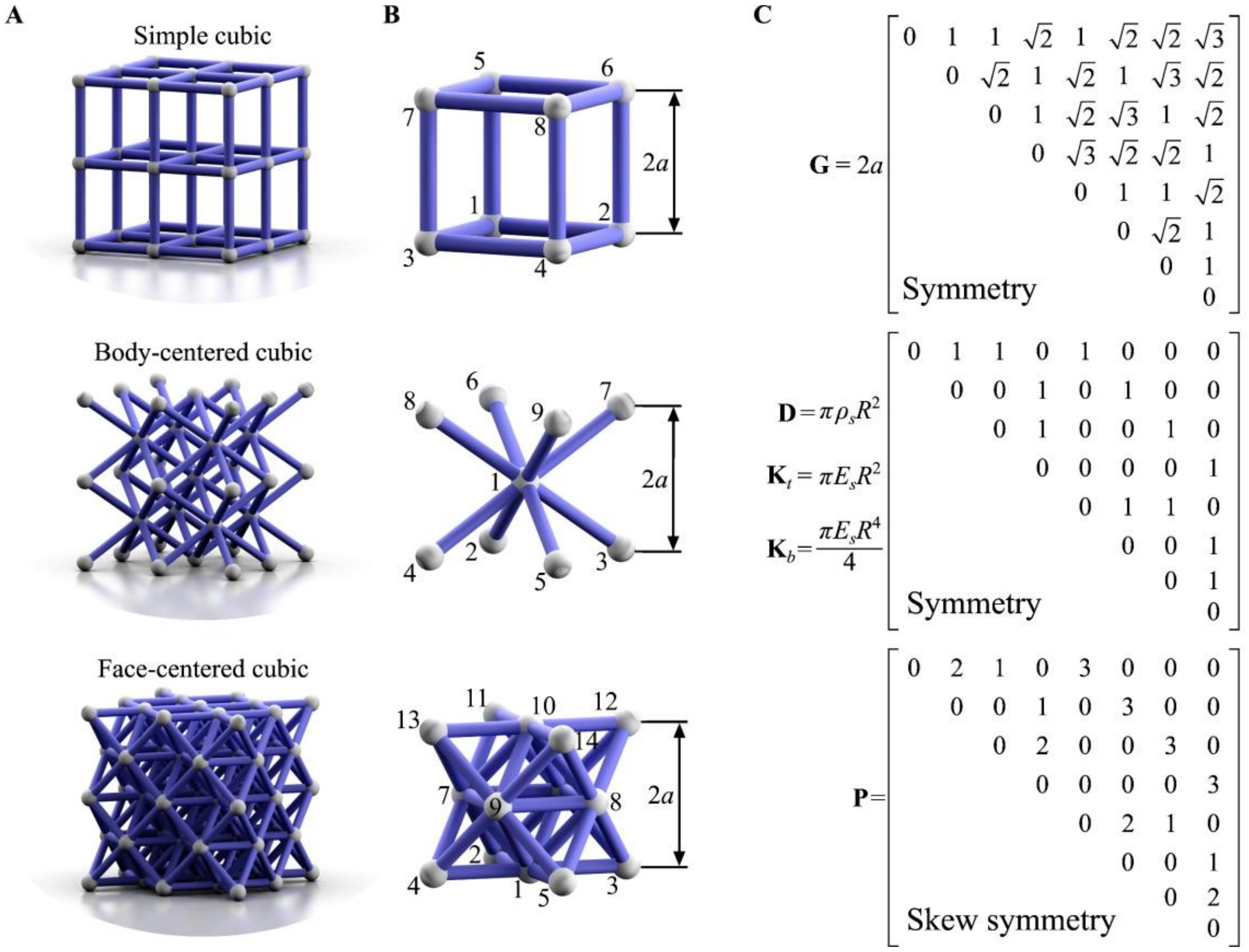

**Fig. 1. Schematic illustration of a series of ad-hoc periodic lattice truss materials. A**, The three lattice truss materials, whose topological configurations are loosely described by the terminologies of simple cubic (SC), body-centered cubic (BCC) and face-centered cubic (FCC) in prior research, comprise a finite packing tessellation with 2×2×2 arrays of the unit cells in 3D space. The white spheres and blue bars indicate the lattice nodes and struts, respectively. **B**, The OBB of the unit cell of these lattices are assumed cubes with a side of length $2a$ and their nodes are randomly numbered. All the struts of these lattice trusses are assumed to have the same cross-section, which is a circular with the diameter of $d$. **C**, The descriptors for the SC (**A**) are calculated based on the geometry and material information. Note that the "Symmetry" indicates that the matrices are symmetric and "Skew symmetry" means that the matrix is skew symmetric and the corresponding entries are omitted. The matrices corresponding to the strut distribution, i.e., $\mathbf{D}$, $\mathbf{K}_t$ and $\mathbf{K}_b$, present the same characteristics of entry distribution except the coefficients, thus allowing the simplified representation.

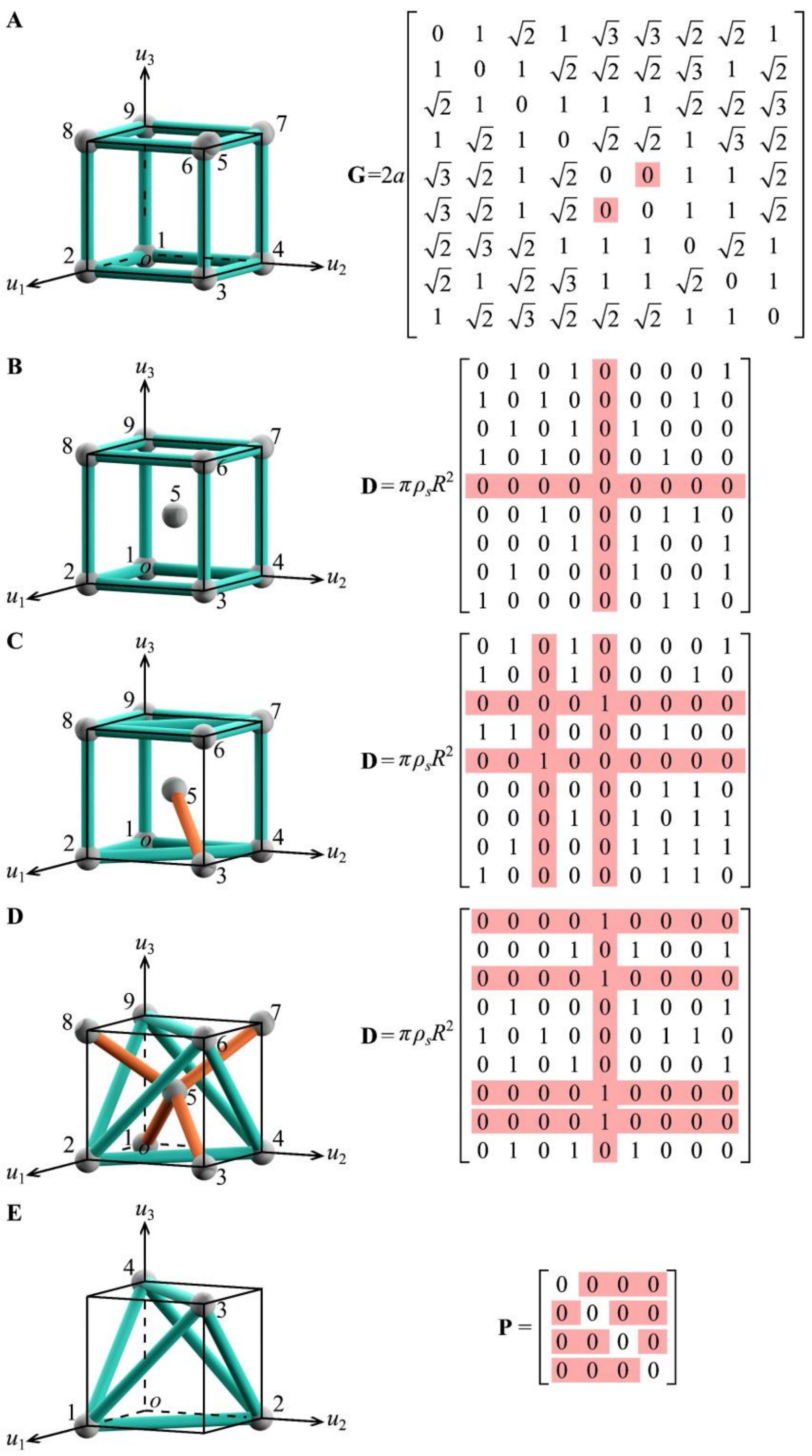

**Fig. 2. The counter-example graphs of the expected characteristics of the PLTMs and the corresponding explicit representations in the matrices. A**, The counter-example to the "No repeated nodes" rule. There are repeated points at one vertex of the unit cell, i.e., *Node* 5 and *Node* 6, between which, the Euclidean distance $r_{56}$ and $r_{65}$ is zero. Correspondingly, two anomalous zeros appear in the geometry matrix. Note that the repeated points are not placed at exactly the same

coordinate for purposes of illustration. **B**, The counter-example to the "No isolated node" rule. An isolated lattice node (*Node* 5) is inserted into a SC cell, generating an all-zero row and column in the density matrix, as marked in red. **C**, The counter-example to the "No isolated strut" rule. The orange strut connecting *Node* 3 and *Node* 5 losses all its links to the rest part of the lattice, causing that all the entries except for $\lambda_{35}$ and $\lambda_{53}$ in two corresponding pairs of rows and columns are zero. **D**, The counter-example to the "No isolated sub-part" rule. In this unit cell, the two sub-parts are formed by placing struts along three non-coplanar edges a tetrahedron (the green part), and by placing struts normal to and at the center of each face of a tetrahedron (the orange part), respectively. From the view of the non-zero entries in the density matrix, *Node* 1, 3, 5, 7 and 8 form a network, in which, *Node* 2, 4, 6 and 9 are missed. Thus, the unit cell is not an intact bulk and violates the "No isolated sub-part" principle. **E**, The counter-example to the "3D periodicity" rule. The tetrahedral unit cell arranged in this unit cell could not be tightly tessellated into 1D, 2D or 3D arrays without any gap along the current principal directions, leading to an all-zero packing matrix.

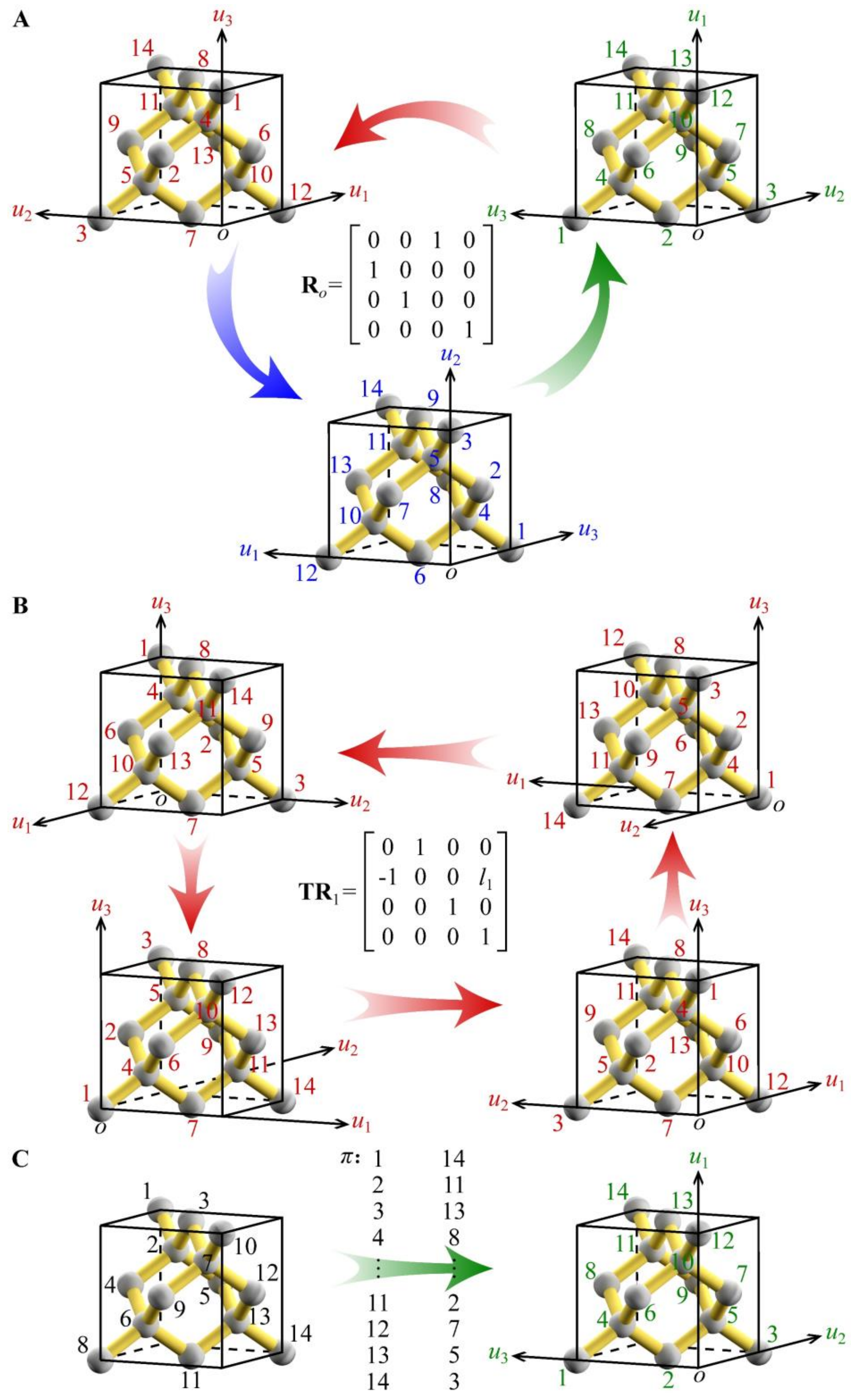

**Fig. 3. The canonical orderings of a lattice unit cell based on an illustrative diagram of the diamond cubic (DC) truss. A**, The mutual transformations of the canonical orderings corresponding to three coordinate systems with the same origin. For instance, three groups of possible Cartesian systems, indicated in red, green and blue respectively, might be applied to the DC when the origin is located at the vertex *o*. The coordinates of the lattice node are computed via

the rotation matrix and the canonical orderings are immediately gained. **B**, If the origin of the coordinate system moves to the vertex on the $u_1$-axis, the new nodes coordinates would be calculated by a rotation and a translation, which can form a translation-rotation matrix $\mathbf{TR}_1$. **C**, For a PLTM with random numbered nodes, a permutation $\pi$ can map the random sequence to the canonical sort and the matrices can be computed in a simple way, instead of starting all over again.

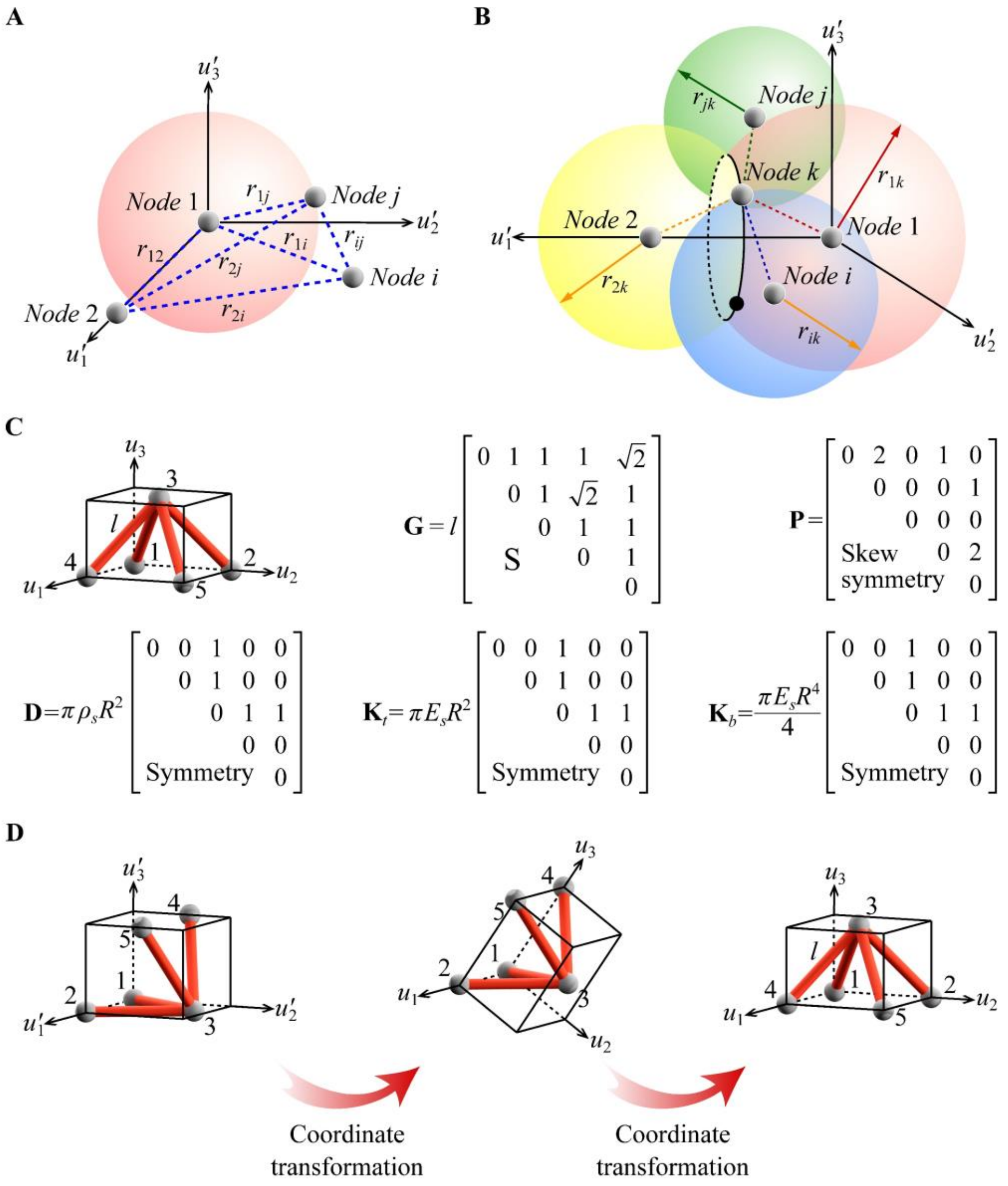

**Fig. 4. Schematic overview of the reconstruction procedure of a PLTM from descriptors. A**, A Cartesian coordinate system is establishes based on the first four non-coplanar lattice nodes. During this step the *Node* 1 and *Node* 2 are used to set the temporary $u_1'$-axis, and the first non-collinear *Node i* is set in the positive direction of the temporary $u_2'$-axis. Then, the right-handed coordinate system can be determined. Two possible positions in the mirror symmetry are present for the first non-coplanar node, perhaps, leading to two mirrored structures. **B**, The coordinates of all the rest lattice nodes can be accurately computed based on the entries in the geometry matrix, followed by generation of lattice struts based on the density matrix, stretching and bending stiffness matrices. **C** and **D** display an example of the reconstruction from given matrices to the structure of a pyramidal unit cell. The length of the struts is assumed to be $l$. The temporary coordinate system might not be the canonical prototype and thus the coordinate transformation should be applied to be consistent with the originally defined one.

# Canonical Descriptors for Periodic Lattice Truss Materials

Ge QI*, Huai-Liang Zheng, Chen-xi Liu, Li MA, Kai-Uwe Schröder

## Supplementary Information

## 1. Examples of descriptors for the ad-hoc PLTMs

The illustrative examples to show the definition of the descriptors are shown in **Fig. 1A**, including SC, BCC and FCC lattice with 2×2×2 arrays of unit cell in space. The numbers of the lattice nodes within one unit cell corresponding to these three types of trusses are 8, 9 and 14, respectively, determining the dimensions of the describing matrices. Once a certain ordering is applied to the lattice nodes within one unit cell of the trusses, as shown in **Fig. 1B**, the entries in the descriptors can be determined, say, by definition.

For the geometry matrix, of course if the space is coordinatized, the Euclidean distance between any nodes can be calculated more easily. In current investigation, it is not necessary to specify this coordinating process since it has no effect on the Euclidean distance. For the density matrix, stretching stiffness matrix and bending stiffness matrix, the entries are gained by literally computing the properties of each strut. In this investigation, all the struts are assumed to be the same except their length. Herein, for these three descriptors, several common coefficients are extracted and the matrices share the same form, in which all the non-zero entries denote 1. All the descriptors corresponding to the periodic lattice truss materials in **Fig. 1** are listed in **Supplementary Table 1~Supplementary Table 3**.

* Corresponding author, E-mail address: qige@ujs.edu.cn (Ge QI)

**Supplementary Table 1 The descriptors for SC lattice.**

| Descriptor | Coefficient | Matrix |
|---|---|---|
| Geometry matrix | $2a$ | $\begin{bmatrix} 0 & 1 & 1 & \sqrt{2} & 1 & \sqrt{2} & \sqrt{2} & \sqrt{3} \\ 1 & 0 & \sqrt{2} & 1 & \sqrt{2} & 1 & \sqrt{3} & \sqrt{2} \\ 1 & \sqrt{2} & 0 & 1 & \sqrt{2} & \sqrt{3} & 1 & \sqrt{2} \\ \sqrt{2} & 1 & 1 & 0 & \sqrt{3} & \sqrt{2} & \sqrt{2} & 1 \\ 1 & \sqrt{2} & \sqrt{2} & \sqrt{3} & 0 & 1 & 1 & \sqrt{2} \\ \sqrt{2} & 1 & \sqrt{3} & \sqrt{2} & 1 & 0 & \sqrt{2} & 1 \\ \sqrt{2} & \sqrt{3} & 1 & \sqrt{2} & 1 & \sqrt{2} & 0 & 1 \\ \sqrt{3} & \sqrt{2} & \sqrt{2} & 1 & \sqrt{2} & 1 & 1 & 0 \end{bmatrix}$ |
| Density matrix | $\pi\rho_s R^2$ | |
| Stretching stiffness matrix | $\pi E_s R^2$ | $\begin{bmatrix} 0 & 1 & 1 & 0 & 1 & 0 & 0 & 0 \\ 1 & 0 & 0 & 1 & 0 & 1 & 0 & 0 \\ 1 & 0 & 0 & 1 & 0 & 0 & 1 & 0 \\ 0 & 1 & 1 & 0 & 0 & 0 & 0 & 1 \\ 1 & 0 & 0 & 0 & 0 & 1 & 1 & 0 \\ 0 & 1 & 0 & 0 & 1 & 0 & 0 & 1 \\ 0 & 0 & 1 & 0 & 1 & 0 & 0 & 1 \\ 0 & 0 & 0 & 1 & 0 & 1 & 1 & 0 \end{bmatrix}$ |
| Bending stiffness matrix | $\frac{\pi E_s R^4}{4}$ | |
| Packing matrix | 1 | $\begin{bmatrix} 0 & 2 & 1 & 0 & 3 & 0 & 0 & 0 \\ -2 & 0 & 0 & 1 & 0 & 3 & 0 & 0 \\ -1 & 0 & 0 & 2 & 0 & 0 & 3 & 0 \\ 0 & -1 & -2 & 0 & 0 & 0 & 0 & 3 \\ -3 & 0 & 0 & 0 & 0 & 2 & 1 & 0 \\ 0 & -3 & 0 & 0 & -2 & 0 & 0 & 1 \\ 0 & 0 & -3 & 0 & -1 & 0 & 0 & 2 \\ 0 & 0 & 0 & -3 & 0 & -1 & -2 & 0 \end{bmatrix}$ |

**Supplementary Table 2 The descriptors for BCC lattice.**

| Descriptor | Coefficient | Matrix |
| --- | --- | --- |
| Geometry matrix | $2a$ | $\begin{bmatrix} 0 & \frac{\sqrt{3}}{2} & \frac{\sqrt{3}}{2} & \frac{\sqrt{3}}{2} & \frac{\sqrt{3}}{2} & \frac{\sqrt{3}}{2} & \frac{\sqrt{3}}{2} & \frac{\sqrt{3}}{2} & \frac{\sqrt{3}}{2} \\ \frac{\sqrt{3}}{2} & 0 & 1 & 1 & \sqrt{2} & 1 & \sqrt{2} & \sqrt{2} & \sqrt{3} \\ \frac{\sqrt{3}}{2} & 1 & 0 & \sqrt{2} & 1 & \sqrt{2} & 1 & \sqrt{3} & \sqrt{2} \\ \frac{\sqrt{3}}{2} & 1 & \sqrt{2} & 0 & 1 & \sqrt{2} & \sqrt{3} & 1 & \sqrt{2} \\ \frac{\sqrt{3}}{2} & \sqrt{2} & 1 & 1 & 0 & \sqrt{3} & \sqrt{2} & \sqrt{2} & 1 \\ \frac{\sqrt{3}}{2} & 1 & \sqrt{2} & \sqrt{2} & \sqrt{3} & 0 & 1 & 1 & \sqrt{2} \\ \frac{\sqrt{3}}{2} & \sqrt{2} & 1 & \sqrt{3} & \sqrt{2} & 1 & 0 & \sqrt{2} & 1 \\ \frac{\sqrt{3}}{2} & \sqrt{2} & \sqrt{3} & 1 & \sqrt{2} & 1 & \sqrt{2} & 0 & 1 \\ \frac{\sqrt{3}}{2} & \sqrt{3} & \sqrt{2} & \sqrt{2} & 1 & \sqrt{2} & 1 & 1 & 0 \end{bmatrix}$ |
| Density matrix | $\pi\rho_s R^2$ | $\begin{bmatrix} 0 & 1 & 1 & 1 & 1 & 1 & 1 & 1 & 1 \\ 1 & 0 & 0 & 0 & 0 & 0 & 0 & 0 & 0 \\ 1 & 0 & 0 & 0 & 0 & 0 & 0 & 0 & 0 \\ 1 & 0 & 0 & 0 & 0 & 0 & 0 & 0 & 0 \\ 1 & 0 & 0 & 0 & 0 & 0 & 0 & 0 & 0 \\ 1 & 0 & 0 & 0 & 0 & 0 & 0 & 0 & 0 \\ 1 & 0 & 0 & 0 & 0 & 0 & 0 & 0 & 0 \\ 1 & 0 & 0 & 0 & 0 & 0 & 0 & 0 & 0 \\ 1 & 0 & 0 & 0 & 0 & 0 & 0 & 0 & 0 \end{bmatrix}$ |
| Stretching stiffness matrix | $\pi E_s R^2$ | |
| Bending stiffness matrix | $\frac{\pi E_s R^4}{4}$ | |
| Packing matrix | 1 | $\begin{bmatrix} 0 & 0 & 0 & 0 & 0 & 0 & 0 & 0 & 0 \\ 0 & 0 & 2 & 1 & 0 & 3 & 0 & 0 & 0 \\ 0 & -2 & 0 & 0 & 1 & 0 & 3 & 0 & 0 \\ 0 & -1 & 0 & 0 & 2 & 0 & 0 & 3 & 0 \\ 0 & 0 & -1 & -2 & 0 & 0 & 0 & 0 & 3 \\ 0 & -3 & 0 & 0 & 0 & 0 & 2 & 1 & 0 \\ 0 & 0 & -3 & 0 & 0 & -2 & 0 & 0 & 1 \\ 0 & 0 & 0 & -3 & 0 & -1 & 0 & 0 & 2 \\ 0 & 0 & 0 & 0 & -3 & 0 & -1 & -2 & 0 \end{bmatrix}$ |

**Supplementary Table 3 The descriptors for FCC lattice.**

| Descriptor | Coefficient | Matrix |
|---|---|---|
| Geometry matrix | $a$ | $\begin{bmatrix} 0 & \sqrt{2} & \sqrt{2} & \sqrt{2} & \sqrt{2} & \sqrt{2} & \sqrt{2} & \sqrt{2} & \sqrt{2} & 2 & \sqrt{6} & \sqrt{6} & \sqrt{6} & \sqrt{6} \\ \sqrt{2} & 0 & 2 & 2 & 2\sqrt{2} & \sqrt{2} & \sqrt{2} & \sqrt{6} & \sqrt{6} & \sqrt{6} & 2 & 2\sqrt{2} & 2\sqrt{2} & 2\sqrt{3} \\ \sqrt{2} & 2 & 0 & 2\sqrt{2} & 2 & \sqrt{2} & \sqrt{6} & \sqrt{2} & \sqrt{6} & \sqrt{6} & 2\sqrt{2} & 2 & 2\sqrt{3} & 2\sqrt{2} \\ \sqrt{2} & 2 & 2\sqrt{2} & 0 & 2 & \sqrt{6} & \sqrt{2} & \sqrt{6} & \sqrt{2} & \sqrt{6} & 2\sqrt{2} & 2\sqrt{3} & 2 & 2\sqrt{2} \\ \sqrt{2} & 2\sqrt{2} & 2 & 2 & 0 & \sqrt{6} & \sqrt{6} & \sqrt{2} & \sqrt{2} & \sqrt{6} & 2\sqrt{3} & 2\sqrt{2} & 2\sqrt{2} & 2 \\ \sqrt{2} & \sqrt{2} & \sqrt{2} & \sqrt{6} & \sqrt{6} & 0 & \sqrt{2} & \sqrt{2} & 2 & \sqrt{2} & \sqrt{2} & \sqrt{2} & \sqrt{6} & \sqrt{6} \\ \sqrt{2} & \sqrt{2} & \sqrt{6} & \sqrt{2} & \sqrt{6} & \sqrt{2} & 0 & 2 & \sqrt{2} & \sqrt{2} & \sqrt{2} & \sqrt{6} & \sqrt{2} & \sqrt{6} \\ \sqrt{2} & \sqrt{6} & \sqrt{2} & \sqrt{6} & \sqrt{2} & \sqrt{2} & 2 & 0 & \sqrt{2} & \sqrt{2} & \sqrt{6} & \sqrt{2} & \sqrt{6} & \sqrt{2} \\ \sqrt{2} & \sqrt{6} & \sqrt{6} & \sqrt{2} & \sqrt{2} & 2 & \sqrt{2} & \sqrt{2} & 0 & \sqrt{2} & \sqrt{6} & \sqrt{6} & \sqrt{2} & \sqrt{2} \\ 2 & \sqrt{6} & \sqrt{6} & \sqrt{6} & \sqrt{6} & \sqrt{2} & \sqrt{2} & \sqrt{2} & \sqrt{2} & 0 & \sqrt{2} & \sqrt{2} & \sqrt{2} & \sqrt{2} \\ \sqrt{6} & 2 & 2\sqrt{2} & 2\sqrt{2} & 2\sqrt{3} & \sqrt{2} & \sqrt{2} & \sqrt{6} & \sqrt{6} & \sqrt{2} & 0 & 2 & 2 & 2\sqrt{2} \\ \sqrt{6} & 2\sqrt{2} & 2 & 2\sqrt{3} & 2\sqrt{2} & \sqrt{2} & \sqrt{6} & \sqrt{2} & \sqrt{6} & \sqrt{2} & 2 & 0 & 2\sqrt{2} & 2 \\ \sqrt{6} & 2\sqrt{2} & 2\sqrt{3} & 2 & 2\sqrt{2} & \sqrt{6} & \sqrt{2} & \sqrt{6} & \sqrt{2} & \sqrt{2} & 2 & 2\sqrt{2} & 0 & 2 \\ \sqrt{6} & 2\sqrt{3} & 2\sqrt{2} & 2\sqrt{2} & 2 & \sqrt{6} & \sqrt{6} & \sqrt{2} & \sqrt{2} & \sqrt{2} & 2\sqrt{2} & 2 & 2 & 0 \end{bmatrix}$ |
| Density matrix | $\pi\rho_s R^2$ | $\begin{bmatrix} 0 & 1 & 1 & 1 & 1 & 1 & 1 & 1 & 1 & 0 & 0 & 0 & 0 & 0 \\ 1 & 0 & 0 & 0 & 0 & 1 & 1 & 0 & 0 & 0 & 0 & 0 & 0 & 0 \\ 1 & 0 & 0 & 0 & 0 & 1 & 0 & 1 & 0 & 0 & 0 & 0 & 0 & 0 \\ 1 & 0 & 0 & 0 & 0 & 0 & 1 & 0 & 1 & 0 & 0 & 0 & 0 & 0 \\ 1 & 0 & 0 & 0 & 0 & 0 & 0 & 1 & 1 & 0 & 0 & 0 & 0 & 0 \\ 1 & 1 & 1 & 0 & 0 & 0 & 1 & 1 & 0 & 1 & 1 & 1 & 0 & 0 \\ 1 & 1 & 0 & 1 & 0 & 1 & 0 & 0 & 1 & 1 & 1 & 0 & 1 & 0 \\ 1 & 0 & 1 & 0 & 1 & 1 & 0 & 0 & 1 & 1 & 0 & 1 & 0 & 1 \\ 1 & 0 & 0 & 1 & 1 & 0 & 1 & 1 & 0 & 1 & 0 & 0 & 1 & 1 \\ 0 & 0 & 0 & 0 & 0 & 1 & 1 & 1 & 1 & 0 & 1 & 1 & 1 & 1 \\ 0 & 0 & 0 & 0 & 0 & 1 & 1 & 0 & 0 & 1 & 0 & 0 & 0 & 0 \\ 0 & 0 & 0 & 0 & 0 & 1 & 0 & 1 & 0 & 1 & 0 & 0 & 0 & 0 \\ 0 & 0 & 0 & 0 & 0 & 0 & 1 & 0 & 1 & 1 & 0 & 0 & 0 & 0 \\ 0 & 0 & 0 & 0 & 0 & 0 & 0 & 1 & 1 & 1 & 0 & 0 & 0 & 0 \end{bmatrix}$ |
| Stretching stiffness matrix | $\pi E_s R^2$ | |
| Bending stiffness matrix | $\frac{\pi E_s R^4}{4}$ | |
| Packing matrix | 1 | $\begin{bmatrix} 0 & 0 & 0 & 0 & 0 & 0 & 0 & 0 & 0 & 3 & 0 & 0 & 0 & 0 \\ 0 & 0 & 2 & 1 & 0 & 0 & 0 & 0 & 0 & 0 & 3 & 0 & 0 & 0 \\ 0 & -2 & 0 & 0 & 1 & 0 & 0 & 0 & 0 & 0 & 0 & 3 & 0 & 0 \\ 0 & -1 & 0 & 0 & 2 & 0 & 0 & 0 & 0 & 0 & 0 & 0 & 3 & 0 \\ 0 & 0 & -1 & -2 & 0 & 0 & 0 & 0 & 0 & 0 & 0 & 0 & 0 & 3 \\ 0 & 0 & 0 & 0 & 0 & 0 & 0 & 0 & 1 & 0 & 0 & 0 & 0 & 0 \\ 0 & 0 & 0 & 0 & 0 & 0 & 0 & 2 & 0 & 0 & 0 & 0 & 0 & 0 \\ 0 & 0 & 0 & 0 & 0 & 0 & -2 & 0 & 0 & 0 & 0 & 0 & 0 & 0 \\ 0 & 0 & 0 & 0 & 0 & -1 & 0 & 0 & 0 & 0 & 0 & 0 & 0 & 0 \\ -3 & 0 & 0 & 0 & 0 & 0 & 0 & 0 & 0 & 0 & 0 & 0 & 0 & 0 \\ 0 & -3 & 0 & 0 & 0 & 0 & 0 & 0 & 0 & 0 & 0 & 2 & 1 & 0 \\ 0 & 0 & -3 & 0 & 0 & 0 & 0 & 0 & 0 & 0 & -2 & 0 & 0 & 1 \\ 0 & 0 & 0 & -3 & 0 & 0 & 0 & 0 & 0 & 0 & -1 & 0 & 0 & 2 \\ 0 & 0 & 0 & 0 & -3 & 0 & 0 & 0 & 0 & 0 & 0 & -1 & -2 & 0 \end{bmatrix}$ |

## 2. An algorithm for identifying strut intersections

The spatial relationships of two struts can naturally be simplified as the arrangement problem of two line segments in 3D space. The idea to determine whether the two struts intersect is quite simple, just requiring the coordinates of the lattice nodes. The process computing the coordinate of each node will be introduced below. Here, provisionally suppose that the space of the unit cell has been coordinated using a canonical Cartesian coordinate system as displayed in **Fig. 3**. Considering, in turn, each pair of two different *struts* $\overline{N_i N_j}$ and $\overline{N_{i'} N_{j'}}$ whose endpoints are *nodes* $N_i$ , $N_j$ , $N_{i'}$ and $N_{j'}$ respectively, there exist ten possible spatial relationships, ranging from collinear struts to skew struts, as shown in **Supplementary Figure 1**. Once one of the cases in **Supplementary Figure 1(C)(D)(H)(I)** is identified for $\overline{N_i N_j}$ and $\overline{N_{i'} N_{j'}}$ , strut intersection occurs and the descriptors are not de-normalized. Here, the spatial relations between two struts can be thought of in the way, finding the *state* $\psi(i, j, i', j')$, whose entry denotes 1 for "strut intersection", 0 for "no intersection" and $\omega$ for "one or both the struts do not exist".

ALGORITHM 3.1

Input: $N_i = [x_{i1}, x_{i2}, x_{i3}]$, $1 \le i \le n$, $\mathbf{G} = [r_{ij}]$, $1 \le i, j \le n$, $\mathbf{D} = [\lambda_{ij}]$, $1 \le i, j \le n$.

Output: $[\psi(i, j, i', j')]$, $1 \le i, j, i', j' \le n$.

1. (Starting with $i, j, i', j' = 1$ ) pick the next $i, j, i', j'$ (ending with $i, j, i', j' = n$).

2. For every $i, j, i', j' = 1, \cdots, n$, if any of $\det\begin{bmatrix} \lambda_{ij} & 0 \\ 0 & \lambda_{i'j'} \end{bmatrix}$, $\det\begin{bmatrix} r_{ii'} & r_{jj'} \\ -r_{jj'} & r_{ii'} \end{bmatrix}$ and $\det\begin{bmatrix} r_{ij'} & r_{i'j} \\ -r_{i'j} & r_{ij'} \end{bmatrix}$ denotes 0, let $\psi(i, j, i', j') = \omega$; otherwise, call $i, j, i', j'$ "good".

3. For every good $i, j, i', j' = 1, \cdots, n$, calculate the equations

$$\begin{cases} a_i x_{i1} + b_i x_{i2} + c_i x_{i3} + d_i = 0 \\ a_j x_{j1} + b_j x_{j2} + c_j x_{j3} + d_j = 0 \end{cases}$$

and

$$\begin{cases} a_{i'} x_{i'1} + b_{i'} x_{i'2} + c_{i'} x_{i'3} + d_{i'} = 0 \\ a_{j'} x_{j'1} + b_{j'} x_{j'2} + c_{j'} x_{j'3} + d_{j'} = 0 \end{cases}.$$

Let

$$\alpha = \begin{bmatrix} a_i & b_i & c_i & d_i \\ a_j & b_j & c_j & d_j \\ a_{i'} & b_{i'} & c_{i'} & d_{i'} \\ a_{j'} & b_{j'} & c_{j'} & d_{j'} \end{bmatrix},$$

$$\beta = \begin{bmatrix} a_i & b_i & c_i \\ a_j & b_j & c_j \\ a_{i'} & b_{i'} & c_{i'} \\ a_{j'} & b_{j'} & c_{j'} \end{bmatrix}.$$

4. For every good $i, j, i', j' = 1, \cdots, n$, if

(a) $rank(\alpha) = 4$, and let $\psi(i, j, i', j') = 0$;

(b) $rank(\alpha) = 3$ and $rank(\beta) = 2$, let $\psi(i, j, i', j') = 0$;

(c) $rank(\alpha) = rank(\beta) = 2$, go to step 5;

(d) $rank(\alpha) = rank(\beta) = 3$, go to step 6.

5. For $rank(\alpha) = rank(\beta) = 2$, if any of

$$\begin{cases} r_{ii'} + r_{i'j} = r_{ij} \\ r_{ii'} r_{i'j} > 0 \end{cases},$$

$$\begin{cases} r_{ij'} + r_{j'j} = r_{ij} \\ r_{ij'} r_{j'j} > 0 \end{cases},$$

$$\begin{cases} r_{i'i} + r_{ij'} = r_{i'j'} \\ r_{i'i} r_{ij'} > 0 \end{cases},$$

is satisfied, let $\psi(i, j, i', j') = 1$; otherwise, let $\psi(i, j, i', j') = 0$. Then go to step 8.

6. For $rank(\alpha) = rank(\beta) = 3$, let

$$\gamma = \begin{bmatrix} x_{j1} - x_{i1} & x_{i'1} - x_{j'1} \\ x_{j2} - x_{i2} & x_{i'2} - x_{j'2} \\ x_{j3} - x_{i3} & x_{i'3} - x_{j'3} \end{bmatrix},$$

$$\chi = \begin{bmatrix} x_{i'1} - x_{i1} \\ x_{i'2} - x_{i2} \\ x_{i'3} - x_{i3} \end{bmatrix},$$

$$\begin{bmatrix} \overline{l}_1 \\ \overline{l}_2 \end{bmatrix} = \left(\gamma^T \gamma\right)^{-1} \left(\gamma^T \chi\right).$$

Then go to step 7.

7. If $0 \le \overline{l}_1 \le 1$, $0 \le \overline{l}_2 \le 1$ and all of $r_{ii'}$, $r_{ij'}$, $r_{ji'}$ as well as $r_{jj'}$ are positive, let $\psi(i, j, i', j') = 1$; otherwise, let $\psi(i, j, i', j') = 0$. Then go to step 8.

8. Return to step 1.

*Analysis*. Step 2 calls $\psi$ "undefined" if the *struts* $\overline{N_i N_j}$ or $\overline{N_{i'} N_{j'}}$ is absent due to the repeated lattice nodes or disconnection, or these two notations indicate the same strut, and designates $N_i$, $N_j$, $N_{i'}$ and $N_{j'}$ as "good" otherwise. Step 3 describes the line in the space which pass through the struts, yielding the coefficient matrix $\beta$ and its augmented matrix $\alpha$. Step 4 preliminarily determines the spatial relations of the two lines into four categories, including coinciding, paralleling, intersecting and skewing, according to the ranks of $\alpha$ and $\beta$. This approach, inherently, discusses the possible number of solutions of the linear equation system corresponding to these lines. Finally, step 5 further distinguishes possible conditions for two collinear struts, while step 6 and step 7 point out that only the case that intersection lies only on the struts, instead of the lines, is identified as strut intersection in this paper. Step 4~step 7 give the $\psi(i, j, i', j')$ for all the good struts. If $\psi(i, j, i', j')$ contains, even if only one, entry with the value of 1, strut intersection occurs and the descriptors are not well set.

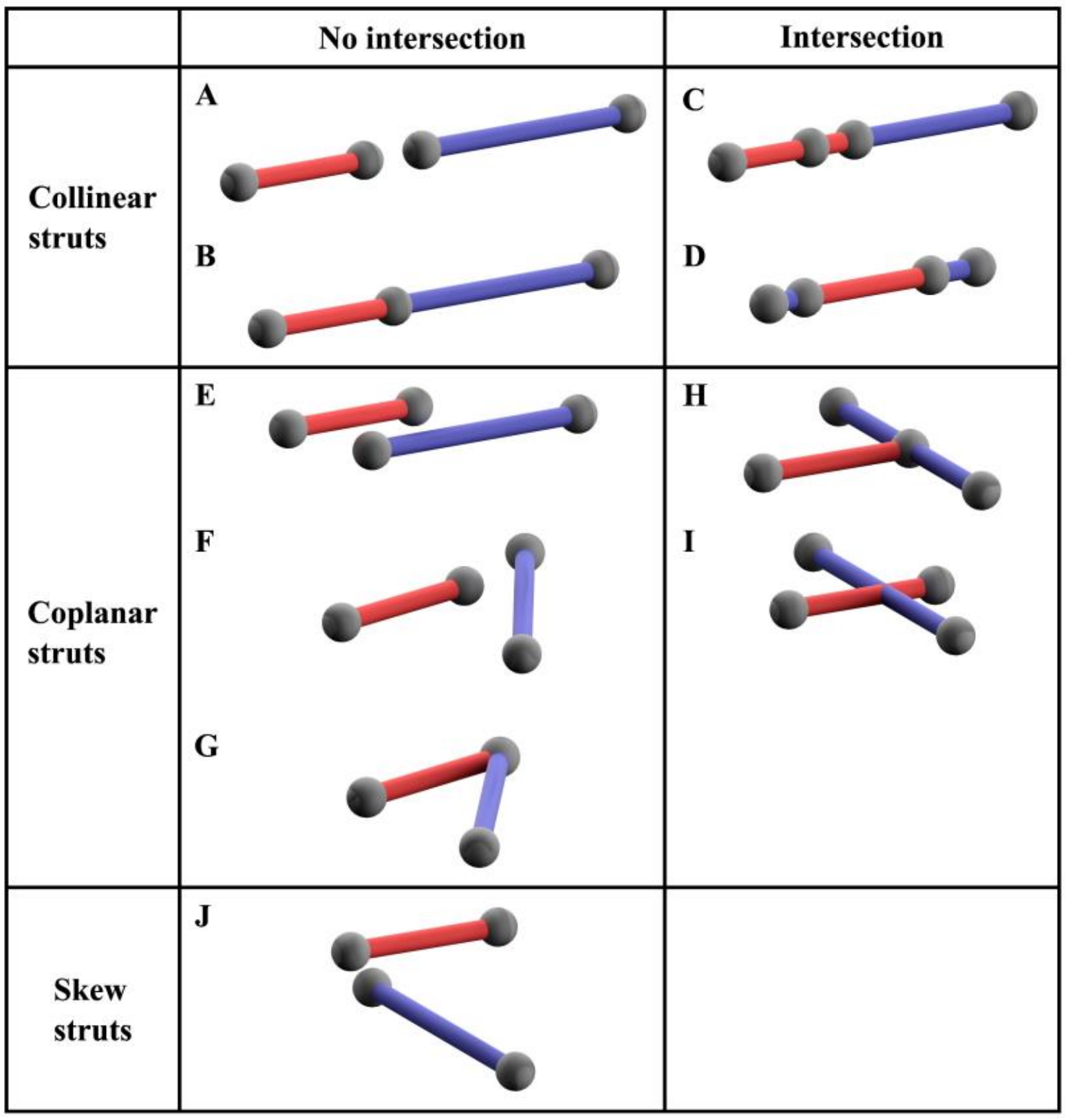

**Supplementary Figure 1 Schematic illustrations of the spatial relationship between two struts. A-C**, The two struts are collinear but (A), they have no any overlapping; (B), they are interconnected only at one node; (C), they overlap partially each other; (D), they overlap fully. **E-I**, The two struts are coplanar (not collinear), but (E), they are parallel; (F), they are not parallel but have no any intersection; (G), they intersect only at one node; (H), an endpoint of one strut lies on the other strut; (I), each strut straddles the line containing the other. **J**, The two struts are skew. The strut intersections are considered for the cases in (C), (D), (H) and (I).

## 3. Examples of all canonical orderings of unit cells

Our definition of the canonical orderings of the unit cells of PLTMs naturally generates 24 and 4 orderings for 3D and 2D lattices, respectively.

For the 3D lattice cells, there are eight possible vertices of the OBB to locate the origin of the Cartesian coordinate system, for each of which, there are three groups of directions to assign the principal axes. **Supplementary Figure 2** shows the examples of all the standard orderings for the DC cell. In each row of this figure, the three orderings, marked in red, green and blue respectively, are obtained based on the same coordinate origin with different principal axis directions.

For the 2D lattice cells, there are four possible vertices of the OBB for the coordinate origin. In contrast to the 3D configurations, only one type of the principal direction for the axes exists for planar lattices, ascribed to the right-handed screw requirement. So, the sorting of lattice nodes can be rigorously resolved as the example of a tetragonal lattice unit cell shown in **Supplementary Figure 3**.

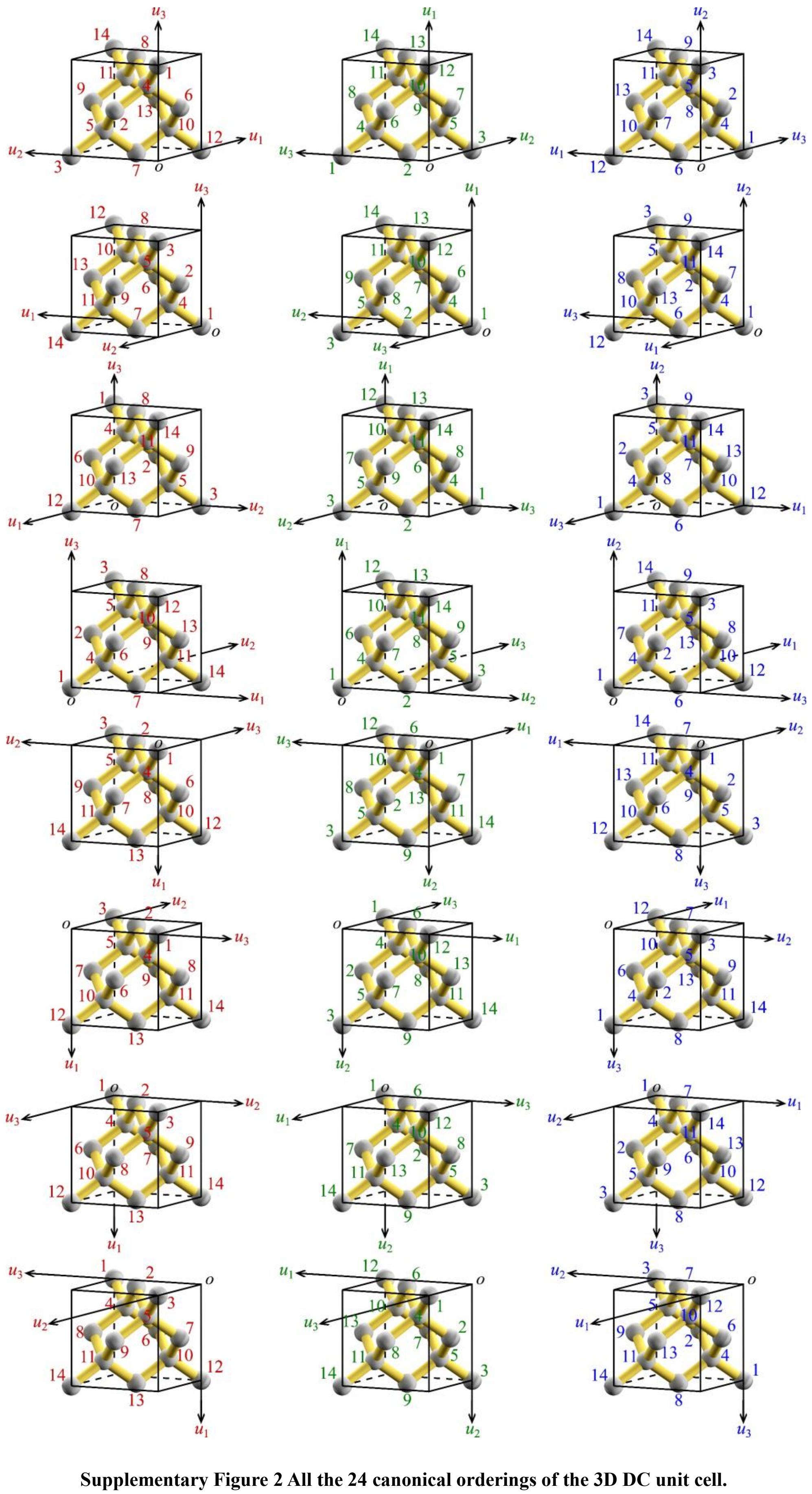

**Supplementary Figure 2 All the 24 canonical orderings of the 3D DC unit cell.**

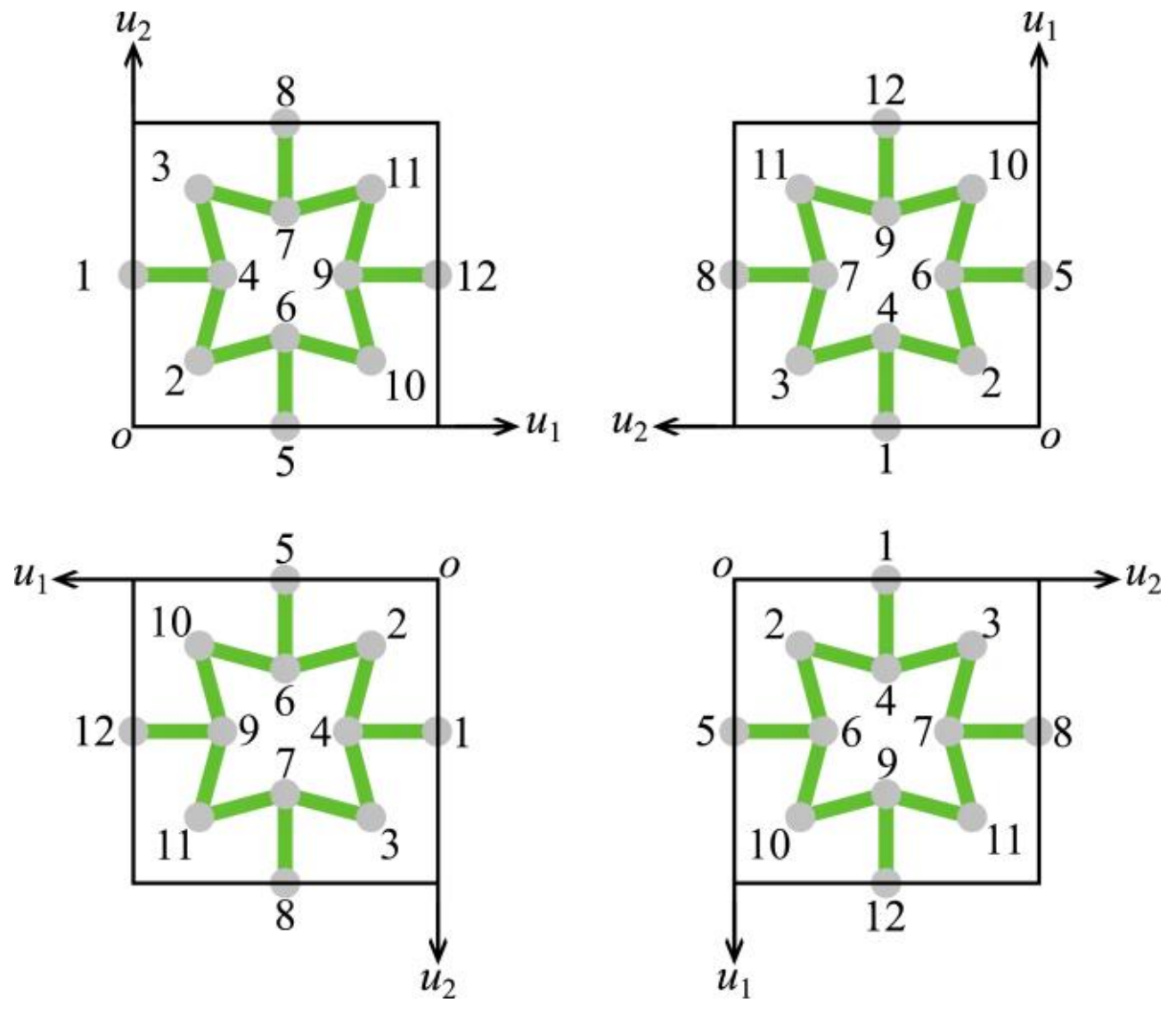

**Supplementary Figure 3 All the 4 canonical orderings of the 2D tetragonal lattice unit cell.**

## 4. Derivation of the element stiffness matrix and equations for PLTMs

When a set of canonical descriptors for a PLTM is given, the information regarding the nodes and struts is specified. Consider a strut connecting *Node i* and *Node j*. The tensile stiffness is $E_{ij}A_{ij}$ according to the stretching stiffness matrix $\mathbf{K}_t$, while the bending stiffness is $E_{ij}I_{ij}$ according to the bending stiffness matrix $\mathbf{K}_b$. And the length of this strut is $r_{ij}$ defined by the geometry matrix. So, the stretching coefficient matrix $\mathbf{C}^{et}$ gives the relationship axial deformation and the axial force

$$\begin{cases} F_i^A = \dfrac{E_{ij}A_{ij}}{r_{ij}}\left(u_i^A - u_j^A\right) \\ F_j^A = \dfrac{E_{ij}A_{ij}}{r_{ij}}\left(-u_i^A + u_j^A\right) \end{cases} \tag{1}$$

where $F_i^A$ and $F_j^A$ denote the axial forces on the two lattice nodes of this strut, while $u_i^A$ and $u_j^A$ denote the axial displacements of the two endpoints of the strut. As for the flexural deformation, the bending coefficient matrices $\mathbf{C}_1^{ef}$, $\mathbf{C}_2^{ef}$ and $\mathbf{C}_3^{ef}$ declare the equations

$$\begin{cases} F_i^S = \dfrac{E_{ij}I_{ij}}{r_{ij}^3}\left(12v_i^S + 6r_{ij}\theta_i - 12v_j^S + 6r_{ij}\theta_j\right) \\ M_i = \dfrac{E_{ij}I_{ij}}{r_{ij}^3}\left(6r_{ij}v_i^S + 4r_{ij}^2\theta_i - 6r_{ij}v_j^S + 2r_{ij}^2\theta_j\right) \\ F_j^S = \dfrac{E_{ij}I_{ij}}{r_{ij}^3}\left(-12v_i^S - 6r_{ij}\theta_i + 12v_j^S - 6r_{ij}\theta_j\right) \\ M_i = \dfrac{E_{ij}I_{ij}}{r_{ij}^3}\left(6r_{ij}v_i^S + 2r_{ij}^2\theta_i - 6r_{ij}v_j^S + 4r_{ij}^2\theta_j\right) \end{cases} \tag{2}$$

If the strut is discretized by *m* beam elements with the same length, we deduce the stiffness equations of each element in local coordinates of the current strut as

$$
\begin{bmatrix} F_{x1}^{eA} \\ F_{y1}^{eS} \\ F_{z1}^{eS} \\ M_{y1}^{e} \\ M_{z1}^{e} \\ F_{x2}^{eA} \\ F_{y2}^{eS} \\ F_{z2}^{eS} \\ M_{y2}^{e} \\ M_{z2}^{e} \end{bmatrix}
= \frac{m^3}{r_{ij}^3}
\begin{bmatrix}
\frac{E_{ij}A_{ij}r_{ij}^2}{m^2} & 0 & 0 & 0 & 0 & \frac{-E_{ij}A_{ij}r_{ij}^2}{m^2} & 0 & 0 & 0 & 0 \\
0 & 12E_{ij}I_{ij} & 0 & 0 & \frac{6E_{ij}I_{ij}r_{ij}}{m} & 0 & -12E_{ij}I_{ij} & 0 & 0 & \frac{6E_{ij}I_{ij}r_{ij}}{m} \\
0 & 0 & 12E_{ij}I_{ij} & \frac{-6E_{ij}I_{ij}r_{ij}}{m} & 0 & 0 & 0 & -12E_{ij}I_{ij} & \frac{-6E_{ij}I_{ij}r_{ij}}{m} & 0 \\
0 & 0 & \frac{-6E_{ij}I_{ij}r_{ij}}{m} & \frac{4E_{ij}I_{ij}r_{ij}^2}{m^2} & 0 & 0 & 0 & \frac{6E_{ij}I_{ij}r_{ij}}{m} & \frac{2E_{ij}I_{ij}r_{ij}^2}{m^2} & 0 \\
0 & \frac{6E_{ij}I_{ij}r_{ij}}{m} & 0 & 0 & \frac{4E_{ij}I_{ij}r_{ij}^2}{m^2} & 0 & \frac{-6E_{ij}I_{ij}r_{ij}}{m} & 0 & 0 & \frac{2E_{ij}I_{ij}r_{ij}^2}{m^2} \\
\frac{-E_{ij}A_{ij}r_{ij}^2}{m^2} & 0 & 0 & 0 & 0 & \frac{E_{ij}A_{ij}r_{ij}^2}{m^2} & 0 & 0 & 0 & 0 \\
0 & -12E_{ij}I_{ij} & 0 & 0 & \frac{-6E_{ij}I_{ij}r_{ij}}{m} & 0 & 12E_{ij}I_{ij} & 0 & 0 & \frac{-6E_{ij}I_{ij}r_{ij}}{m} \\
0 & 0 & -12E_{ij}I_{ij} & \frac{6E_{ij}I_{ij}r_{ij}}{m} & 0 & 0 & 0 & 12E_{ij}I_{ij} & \frac{6E_{ij}I_{ij}r_{ij}}{m} & 0 \\
0 & 0 & \frac{-6E_{ij}I_{ij}r_{ij}}{m} & \frac{2E_{ij}I_{ij}r_{ij}^2}{m^2} & 0 & 0 & 0 & \frac{6E_{ij}I_{ij}r_{ij}}{m} & \frac{4E_{ij}I_{ij}r_{ij}^2}{m^2} & 0 \\
0 & \frac{6E_{ij}I_{ij}r_{ij}}{m} & 0 & 0 & \frac{2E_{ij}I_{ij}r_{ij}^2}{m^2} & 0 & \frac{-6E_{ij}I_{ij}r_{ij}}{m} & 0 & 0 & \frac{4E_{ij}I_{ij}r_{ij}^2}{m^2}
\end{bmatrix}
\begin{bmatrix} u_1 \\ v_1 \\ w_1 \\ \theta_{y1} \\ \theta_{z1} \\ u_2 \\ v_2 \\ w_2 \\ \theta_{y2} \\ \theta_{z2} \end{bmatrix} \quad (3)
$$

The following steps, ranging from assembling the global stiffness matrix and coordinate transformation, have widely been used in FEA, thus they are not covered in this paper.